\documentclass[aps,prb,reprint,superscriptaddress,amsmath,amssymb,longbibliography]{revtex4-2}

\usepackage{graphicx}
\usepackage{bm}
\usepackage{booktabs}
\usepackage{array}[=2016-10-06]
\usepackage{hyperref}
\hypersetup{colorlinks=true,linkcolor=blue,citecolor=blue,urlcolor=blue}
\graphicspath{{figures/}}
\newcommand{\unitEFG}{\mathrm{V}/\text{\AA}^{2}}
\makeatletter
\let\prb@originaldoauthor\doauthor
\def\doauthor#1#2#3{%
  \def\prb@currentname{#1}\def\prb@sharedname{Lin Li}%
  \ifx\prb@currentname\prb@sharedname
    \ignorespaces#1\unskip\@listcomma
    \begingroup#3\endgroup{\comma@space}{}%
    \frontmatter@footnotemark{1}% Reuse the first author's shared asterisk.
    \textsuperscript{,}%
    \frontmatter@footnote{Contact author: \href{mailto:lilin1@siom.ac.cn}{lilin1@siom.ac.cn}}%
    \space\@listand
  \else
    \prb@originaldoauthor{#1}{#2}{#3}%
  \fi
}
\makeatother

\begin{document}

\title{Th$^{4+}$ ion cluster structures and electric field gradients in $^{229}$Th:CaF$_2$}

\author{Yuxiang Li}
\thanks{These authors contributed equally to this work.}
\affiliation{Research Center of Laser Crystal, Shanghai Institute of Optics and Fine Mechanics, Chinese Academy of Sciences, Shanghai 201800, China.}
\affiliation{Center of Materials Science and Optoelectronics Engineering, University of Chinese Academy of Sciences, Beijing 100049, China.}
\author{Lin Li}
\thanks{These authors contributed equally to this work.}
\email[Contact author: ]{lilin1@siom.ac.cn}
\affiliation{Research Center of Laser Crystal, Shanghai Institute of Optics and Fine Mechanics, Chinese Academy of Sciences, Shanghai 201800, China.}
\affiliation{Center of Materials Science and Optoelectronics Engineering, University of Chinese Academy of Sciences, Beijing 100049, China.}
\author{Qiaorui Gong}
\affiliation{Research Center of Laser Crystal, Shanghai Institute of Optics and Fine Mechanics, Chinese Academy of Sciences, Shanghai 201800, China.}
\affiliation{Center of Materials Science and Optoelectronics Engineering, University of Chinese Academy of Sciences, Beijing 100049, China.}
\author{Shanming Li}
\affiliation{Research Center of Laser Crystal, Shanghai Institute of Optics and Fine Mechanics, Chinese Academy of Sciences, Shanghai 201800, China.}
\affiliation{Center of Materials Science and Optoelectronics Engineering, University of Chinese Academy of Sciences, Beijing 100049, China.}
\author{Chengchun Zhao}
\affiliation{Research Center of Laser Crystal, Shanghai Institute of Optics and Fine Mechanics, Chinese Academy of Sciences, Shanghai 201800, China.}
\affiliation{Center of Materials Science and Optoelectronics Engineering, University of Chinese Academy of Sciences, Beijing 100049, China.}
\author{Lu Deng}
\affiliation{Advanced Laser and Optoelectronic Functional Materials Department, Shanghai Institute of Optics and Fine Mechanics, Chinese Academy of Sciences, Shanghai 201800, China.}
\author{Peixiong Zhang}
\email[Contact author: ]{pxzhang@jnu.edu.cn}
\affiliation{Research Center of Laser Crystal, Shanghai Institute of Optics and Fine Mechanics, Chinese Academy of Sciences, Shanghai 201800, China.}
\affiliation{Department of Optoelectronic Engineering, Jinan University, Guangzhou, Guangdong 510632, China.}
\author{Yin Hang}
\email[Contact author: ]{yhang@siom.ac.cn}
\affiliation{Research Center of Laser Crystal, Shanghai Institute of Optics and Fine Mechanics, Chinese Academy of Sciences, Shanghai 201800, China.}
\affiliation{Center of Materials Science and Optoelectronics Engineering, University of Chinese Academy of Sciences, Beijing 100049, China.}
\author{Longsheng Ma}
\affiliation{State Key Laboratory of Precision Spectroscopy, East China Normal University, Shanghai 200062, China.}
\author{Shining Zhu}
\affiliation{National Laboratory of Solid-State Microstructures, School of Physics, Collaborative Innovation Center of Advanced Microstructures, Nanjing University, Nanjing, Jiangsu 210093, China.}

\begin{abstract}

The low-energy nuclear isomeric transition of $^{229}$Th is a promising frequency reference for a solid-state nuclear optical clock. Wide-band-gap fluoride crystals such as CaF$_2$ offer transparent hosts for embedding Th nuclei at high density. Substitution of Th$^{4+}$ for Ca$^{2+}$ requires charge compensation, linking the local defect structure to nuclear quadrupole splitting. We use first-principles calculations to examine neutral Th dimers compensated by four interstitial F ions ($\mathrm{F}_{i}^{-}$). Screening different compensation geometries by their relative formation energies identifies a family of low-energy configurations with closely similar local structures. These configurations share large electric field gradients and small asymmetry parameters, consistent with the experimental high-EFG site 3. Their calculated quadrupole splittings are further compared with the measured C13 spectra. Together, the energetic and spectroscopic results support $\mathrm{F}_{i}^{-}$-compensated Th dimers as candidate microscopic structures for site 3 in Th:CaF$_2$.

\end{abstract}

\maketitle

\section{Introduction}

Optical clocks are among the frequency references with the lowest uncertainties and highest stability in current frequency metrology. They have also become important platforms for testing fundamental physics, including searches for dark-matter candidates and temporal variation of fundamental constants \cite{ref01,ref02}. Conventional optical atomic clocks use electronic-level transitions as frequency references and have reached very high accuracy and stability, but they still require precise control of electromagnetic environmental effects such as Stark shifts, Zeeman shifts, blackbody-radiation shifts, and optical-lattice perturbations. $^{229}$Th  possesses a unique, laser-accessible, extremely low-energy excited state (8.356 eV), which can be excited by a 148.38-nm vacuum ultraviolet (VUV) laser, providing an important basis for the development of next-generation optical clocks \cite{ref03,ref04,ref05,ref06,ref07,ref08,ref09}.

The research approaches for the $^{229}$Th nuclear clock encompass single-ion and solid-state nuclear clocks \cite{ref10,ref11,ref12}. The potential advantage of a solid-state nuclear optical clock is that a transparent crystal can host a macroscopic number of $^{229}$Th nuclei in a compact, mechanically stable system, potentially enhancing the collective detection strength and averaging efficiency of the vacuum-ultraviolet (VUV) nuclear signal. CaF$_2$ crystal is among the most extensively studied $^{229}$Th host materials for this purpose because of its wide band gap, VUV transparency, relatively simple fluorite structure, and established basis for growing Th-doped single crystals \cite{ref13,ref14,ref15,ref16,ref17}. Studies of radiative decay, direct laser excitation, frequency-ratio measurements, temperature sensitivity, frequency reproducibility, laser-induced quenching, and site-resolved spectroscopy indicate that solid-state $^{229}$Th nuclear spectroscopy has progressed from proof-of-concept demonstrations to an experimental stage in which the microscopic materials environment is a limiting factor \cite{ref18,ref19,ref20,ref21,ref22,ref23,ref24,ref25,ref26}. Recent studies demonstrated continuous-wave VUV laser interrogation of the 148.4 nm $^{229}$Th nuclear transition in crystalline hosts and nuclear-clock operation with closed-loop frequency stabilization, representing an important step toward solid-state nuclear-clock operation \cite{ref27,ref28}.

Current solid-state $^{229}$Th spectra resolve a near-zero-EFG O-like/site 1 center, a D-center/site 2 with $V_{zz}$ of approximately 100 V/\AA$^2$ and $\eta$ $\approx$ 0.6, and high-EFG site 3/site 4 centers with $V_{zz}$ of $\sim$260--320 V/\AA$^2$ \cite{ref20,ref21,ref25,ref26}. These centers correspond to distinct local defect environments and quadrupole-splitting parameters. If their relative weights, frequency shifts, and temperature responses cannot be experimentally separated, spectral fitting is affected by center-selection bias and inhomogeneous broadening, reducing the precision of nuclear-transition center-frequency extraction, frequency reproducibility, and frequency-locking stability \cite{ref16,ref17,ref23,ref24}. Furthermore, in Th-doped crystals, increasing Th concentration to enhance the number of nuclei offers a signal advantage but also increases material complexity. In practical Th-doped crystals, charge compensation and lattice relaxation lead to multiple non-equivalent local configurations. These structures can affect the spectra through at least two classes of local electronic-structure descriptors: the electron density and shielding at the Th nucleus are related to the isomer shift or chemical shift, whereas the EFG tensor characterizes the anisotropy of the local electrostatic potential and modifies the quadrupole splitting of nuclear energy levels through the nuclear electric-quadrupole interaction \cite{ref29,ref30}. Thus, the EFG is not a complete proxy for the center frequency or linewidth of a nuclear optical clock, but it is one of the key physical quantities for identifying non-equivalent Th sites, linking microscopic defect environments, assigning split spectral lines, and understanding spectral precision degradation \cite{ref21,ref22,ref23,ref24,ref25,ref26}.

The local defect environment of Th is first constrained by charge compensation. X-ray absorption fine-structure measurements support substitutional Th$^{4+}$ with neighboring interstitial-F compensation in CaF$_2$ \cite{ref31}. This model is also supported by electronic-structure studies \cite{ref13,ref32,ref33}. There is ample theoretical and experimental evidence that aliovalent cation doping in fluorite-structured materials readily induces defect cluster aggregates \cite{ref35,ref36,ref37,ref38}. For Th:CaF$_2$, a recent site-resolved laser Mössbauer study further used HAADF-STEM imaging of an approximately 20 nm-thick high-concentration sample and reported that Th-containing atomic columns usually show aggregated distributions. A nearest-neighbor Th--Th dimer DFT model without local charge compensation gave a D-like candidate assignment with $V_{zz}$ of about 95 V/\AA$^2$ and $\eta$ of about 0.6 \cite{ref26}. Research on defect clusters in rare-earth-ion-doped fluorite-structured fluorides further confirms the correlation between local structure and spectroscopy \cite{ref39,ref40,ref53}. The defect structure affects the Th--F coordination, local electronic structure, and EFG of Th sites \cite{ref32,ref33}. However, spectra of $^{229}$Th:CaF$_2$ simultaneously include near-zero-EFG centers, D-like centers of $\sim$100 V/\AA$^2$, and high-EFG centers of $\sim$260--320 V/\AA$^2$ \cite{ref20,ref21,ref25,ref26}. Therefore, it is necessary to compare interstitial-F-compensated and Ca-vacancy-compensated monomer clusters with dimer clusters under the same computational framework and to correlate the relaxed structures, relative static energy, and EFG tensor. This work investigated charge-balanced Th$^{4+}$ defect configurations in CaF$_2$ and analyzed the effects of charge-compensated geometry on local coordination, lattice relaxation, defect-defect interactions, and the EFG tensor descriptors relevant to experimental nuclear spectra.

\section{Methods}

\subsection{Construction of defect models}

All defect structures were constructed from a (2 $\times$ 2 $\times$ 2) fluorite CaF$_2$ supercell. The initial cubic supercell lattice parameter was 10.8046 \AA, and the cell shape and volume were subsequently allowed to relax during structural optimization. In fluorite CaF$_2$, substitution of Th on a Ca site can be written in Kröger--Vink notation as $\mathrm{Th}_{\mathrm{Ca}}^{\bullet\bullet}$. Charge neutrality can be achieved by two interstitial $\mathrm{F}_{i}^{-}$ ions, one Ca$^{2+}$ vacancy $V_{\mathrm{Ca}}^{\prime\prime}$, or more complex multi-defect clusters. Th was introduced by replacing Ca atoms on the Ca sublattice and is denoted as $\mathrm{Th}_{\mathrm{Ca}}^{\bullet\bullet}$. The $\mathrm{F}_{i}^{-}$ monomers have the composition Ca$_{31}$ThF$_{66}$, corresponding to $\mathrm{Th}_{\mathrm{Ca}}^{\bullet\bullet}$ + $2\mathrm{F}_{i}^{-}$. $\mathrm{F}_{i}^{-}$ sites were selected from polyhedral void positions. The $\mathrm{F}_{i}^{-}$ dimers have the composition Ca$_{30}$Th$_2$F$_{68}$, corresponding to 2$\mathrm{Th}_{\mathrm{Ca}}^{\bullet\bullet}$ + $4\mathrm{F}_{i}^{-}$. Ca-vacancy compensation models were also included as competing compensation mechanisms. The $V_{\mathrm{Ca}}$ monomer has the composition Ca$_{30}$ThF$_{64}$, corresponding to $\mathrm{Th}_{\mathrm{Ca}}^{\bullet\bullet}$ + $V_{\mathrm{Ca}}^{\prime\prime}$, with the Ca vacancy placed in different coordination shells around Th. The Ca-vacancy-compensated dimer model has the composition Ca$_{28}$Th$_2$F$_{64}$, corresponding to 2$\mathrm{Th}_{\mathrm{Ca}}^{\bullet\bullet}$ + 2$V_{\mathrm{Ca}}^{\prime\prime}$. These structures were used to analyze Th-site equivalence, local compensation differences, and the EFG distribution range induced by clustering.

\subsection{Structural optimization}

Structural optimizations were performed with VASP \cite{ref41,ref42}, using the generalized gradient approximation of Perdew, Burke, and Ernzerhof (PBE) \cite{ref43} and projector augmented-wave (PAW) potentials \cite{ref44,ref45}. The screening series used Ca\_sv--Th--F potentials, a 650 eV plane-wave cutoff for relaxation, and a $\Gamma$-centered 3 $\times$ 3 $\times$ 3 k-point mesh. Atomic positions and lattice parameters were relaxed without spin polarization or symmetry constraints, with a force threshold of 0.02 eV/\AA. Static EFG calculations used 700 eV and an electronic energy threshold of $10^{-8}$ eV, retaining the mesh and potentials. This screening series is scalar relativistic and does not include spin--orbit coupling (SOC). Separate cutoff, k-point, and lattice-constraint controls are described in Secs. I--III and VI of the Supplemental Material \cite{ref46}. Because some controls use different PAW datasets and charged geometries, they do not by themselves establish numerical convergence of every neutral dimer.

\subsection{Relative defect energies}

For configurations with identical composition and charge state, chemical-potential contributions cancel in relative formation energies \cite{ref47,ref48,ref49}. We compare the static total energies of the neutral Ca$_{30}$Th$_2$F$_{68}$ dimer family using

\begin{equation}
\Delta E_{\mathrm{rel},i}=E_{\mathrm{tot}}(C_i)-E_{\mathrm{tot}}(C_{01}).
\label{eq:relative}\end{equation}

Here $C_i$ denotes dimer configuration i and $C_{01}$ is the lowest-energy sampled configuration. All configurations in this comparison use the same composition, charge state, PAW dataset, cutoff, and k-point mesh. The reference-independent energy differences describe the sampled dimer family; they do not determine aggregation energies relative to isolated defects or equilibrium defect populations. Differences of a few meV are not treated as a robust ordering of distinct species. Absolute formation energies of charged defects would additionally require chemical potentials, a Fermi-level reference, and appropriate charged-cell corrections \cite{ref50,ref51,ref52}, which are outside the scope of this comparison.

\subsection{Static EFG calculations and nuclear quadrupole splitting}

Static calculations on the relaxed structures yielded Th-site EFG tensors, whose principal components obey $|V_{zz}|$ $\geq$ $|V_{yy}|$ $\geq$ $|V_{xx}|$. The asymmetry is $\eta$ = ($V_{xx}$ - $V_{yy}$)/$V_{zz}$, with EFGs reported in V/\AA$^2$. These parameters are compared with experiment and used to model the C13 spectra in Fig. 4. EFG conventions, static calculation settings, and spectral modeling are detailed in Secs. II, III, and VII of the Supplemental Material \cite{ref46}, respectively.

The EFG splits the ground state ($I_g$= 5/2) and isomeric state ($I_e$ = 3/2) of $^{229}$Th through the nuclear electric-quadrupole interaction. In the EFG principal-axis frame, the Hamiltonian for nuclear state $s\in\{g,e\}$, expressed using dimensionless spin operators, is:

\begin{equation}
\begin{split}
\hat H_Q^{(s)}={}&\frac{Q_sV_{zz}}{4I_s(2I_s-1)}\\
&\times\left[3\hat I_z^2-I_s(I_s+1)+\frac{\eta}{2}(\hat I_+^2+\hat I_-^2)\right].
\end{split}
\label{eq:hamiltonian}\end{equation}

Here $Q_s$ is the laboratory-frame spectroscopic electric quadrupole moment, including the unit of charge, and must not be replaced by the intrinsic quadrupole moment $Q_0$. We use $Q_g$ = 3.11(2) eb as a reference value \cite{ref56}; $Q_e$ is approximately 1.77 eb when the measured ratio $Q_e$/$Q_g$ is used \cite{ref20}. For the C13 fits, $Q_g$ is fixed at 3.11 eb and $Q_e$/$Q_g$ is fitted within 0.53--0.61, consistently with Sec. VII of the Supplemental Material \cite{ref46}. The 6 $\times$ 6 ground-state and 4 $\times$ 4 isomeric-state Hamiltonians are diagonalized numerically to obtain $E_{g,k}$ and $E_{e,l}$, respectively. The quadrupole shift of each transition relative to the unsplit frequency $\nu_0$ is

\begin{equation}
\nu_{lk}=\nu_0+\Delta\nu_{lk},\qquad
\Delta\nu_{lk}=\frac{E_{e,l}-E_{g,k}}{h}.
\label{eq:frequency}\end{equation}

For $\eta$ $\ne$ 0, components with different magnetic quantum numbers mix, and the line positions are determined from the numerical eigenvalues. With $Q_s$ expressed in eb and $V_{zz}$ in V/\AA$^2$ (1 V/\AA$^2$ = $10^{20}$ V/m$^2$), the basic coupling scale is converted as

\begin{equation}
\frac{Q_s V_{zz}}{h}=2.417989242\,\mathrm{MHz}
\left(\frac{Q_s}{e\,\mathrm b}\right)
\left(\frac{V_{zz}}{\unitEFG}\right).
\label{eq:units}\end{equation}

\section{Results and discussion}

\subsection{Th$^{4+}$ clusters in Th:CaF$_2$}

HAADF-STEM observations provide evidence for Th aggregation in a high-concentration Th:CaF$_2$ sample \cite{ref26}. Direct imaging of rare-earth clusters has also been reported for Yb:CaF$_2$ \cite{ref54}, but that observation does not identify the structure or abundance of Th clusters. When Th$^{4+}$ substitutes for a Ca$^{2+}$ site, its effective excess charge can be compensated by two F interstitials or one Ca vacancy. Interstitial-F compensation near Th is supported by structural and theoretical studies \cite{ref13,ref31,ref32,ref33}. Within the formation-energy convention of Ref. \cite{ref17}, the 90$^\circ$ and 180$^\circ$ interstitial-F configurations have reported energies of -6.68 and -5.87 eV, respectively. This comparison motivates examining the lower-energy 90$^\circ$ starting motif in Fig. 1; it does not determine its equilibrium population. Relaxation changes the local Th--F distances and expands the finite supercell containing one [Th--F$_{10}$]$^{6-}$ unit. This model corresponds to one Th atom per 32 cation sites, or 3.125\%, and its lattice response should not be extrapolated directly to the dilute limit.

\begin{figure}[!hb]\centering
\includegraphics[width=\columnwidth]{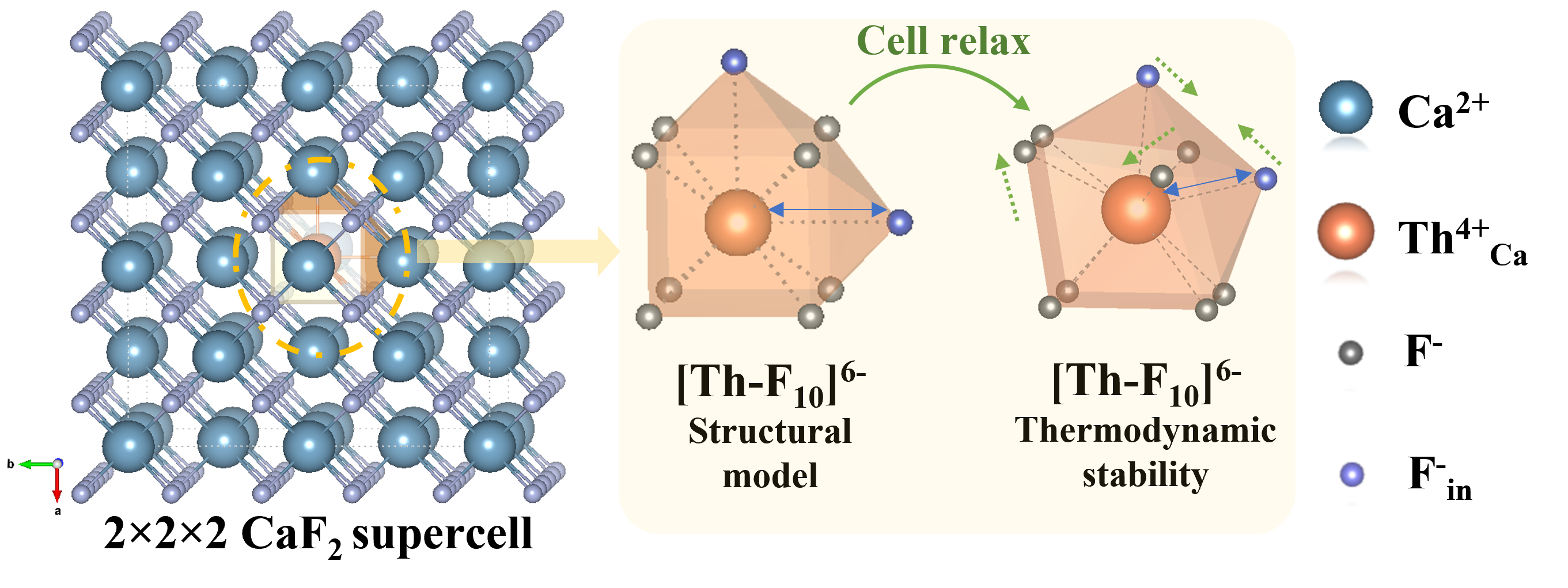}
\caption{Crystal structure of Th-doped CaF$_2$ and the [Th--F$_{10}$]$^{6-}$ monomer cluster.}\label{fig:1}
\end{figure}

\begin{figure*}[!t]\centering
\includegraphics[width=\textwidth]{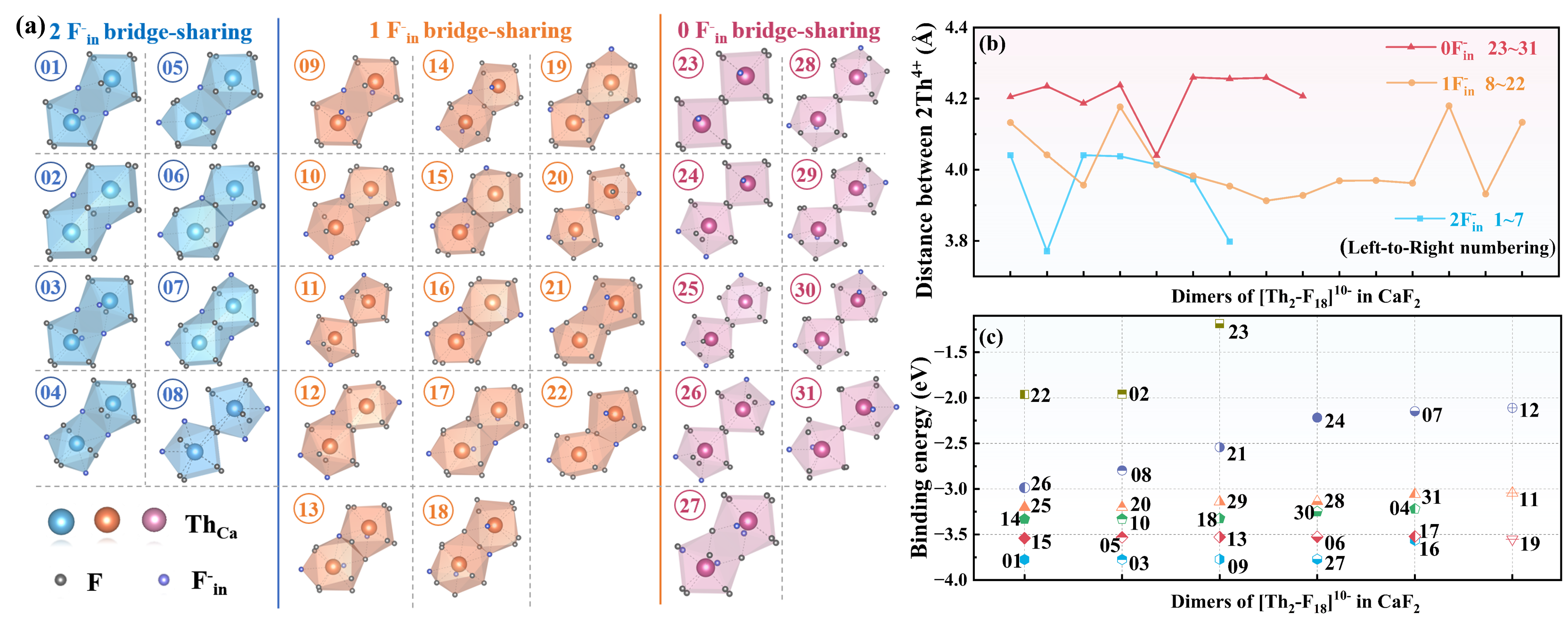}
\caption{Th$^{4+}$ dimer-cluster configurations in CaF$_2$. (a) Relaxed structures. (b) Distances between two Th$^{4+}$ ions. (c) Binding energies.}\label{fig:2}
\end{figure*}

The neutral dimer family contains two substitutional Th atoms and four compensating F interstitials. Thirty-one initial interstitial arrangements were sampled (Fig. 2). Their relaxed Th--Th distances depend on the bridge-sharing geometry, reflecting changes in local coordination and lattice distortion. Within this fixed-composition family, the relative static energies span 0--2.587379 eV. Configurations 02, 07, 12, and 22--24 lie well above the lowest sampled family, whereas 01, 03, 09, and 27 are separated by less than 0.004 eV. These comparisons characterize the sampled structures; they do not determine aggregation energies or defect populations under growth conditions.

In our previous study, the VUV transmittance near 150 nm decreased from approximately 62\% to 53\% as the $^{232}$Th concentration increased from 1.91 $\times$ $10^{20}$ to 2.76 $\times$ $10^{20}$ cm$^{-3}$ \cite{ref17}. Defect-related absorption, scattering, and local structural disorder may contribute to this trend \cite{ref14,ref15,ref16,ref17}. Transmission measurements alone do not demonstrate Th dimer formation. Establishing a causal link between clustering, VUV loss, and nuclear spectral inhomogeneity requires correlated structural and spectroscopic measurements on the same samples \cite{ref21,ref22,ref25,ref26}.

\subsection{Comparison of EFG results with experiment}

The distinctive focus of this work is the relationship between local charge compensation, dimer geometry, and nuclear spectroscopic signatures within a charge-neutral Th-dimer family. We systematically compare nearest-neighbor Th pairs compensated by four interstitial F$^{-}$ ions at fixed composition and charge. Combining relative static energies, site-resolved EFGs, and quadrupole-splitting calculations allows us to assess candidate dimer structures for the high-EFG spectral center. In particular, the similar relative formation energies, relaxed geometries, EFGs, and $\eta$ values of configurations 01, 03, 09, and 27 suggest closely related realizations of a common local structural motif, whose small site-specific differences can nevertheless produce different quadrupole line positions. By comparison, Rehn et al. \cite{ref55} favor an isolated Th--2$\mathrm{F}_{i}^{-}$ defect for the dominant quadrupole-split clock site; their PBE/SOC 90$^\circ$ motif yields $V_{zz}$ = 107.650 V/\AA$^2$ and $\eta$ = 0.173. The locally uncompensated Th pair in their association analysis has charge +4, whereas our two-Th/four-interstitial-F model is neutral. Our analysis therefore addresses a distinct compensation mechanism and its possible contribution to the multiple nuclear environments in Th:CaF$_2$. These differences in charge state and local compensation must be taken into account when comparing the calculated EFGs.

Structural relaxation from different initial interstitial-F arrangements yields similar local structures among the lowest-energy dimer configurations. Based on the screening comparison in Sec. III A, configurations 01, 03, 09, and 27 have low relative static energies and similar structures. Therefore, EFG calculations were further performed for these configurations. The experimental centers in Table I provide benchmarks for this comparison. We use D-like for the combination $|V_{zz}|$ $\approx$ 100--110 V/\AA$^2$ and $\eta$ $\approx$ 0.6, while retaining the literature labels D-center and site 2 without assuming identical microscopic structures \cite{ref20,ref21,ref25,ref26}. The site-resolved spectra also contain an O-like, near-zero-EFG center and high-EFG centers near 260 and 320 V/\AA$^2$ \cite{ref26}.  Both $|V_{zz}|$ and $\eta$, the relation between the two Th sites, and the energy ordering within the dimer family enter this assessment.

The lowest-energy dimer family has a high-EFG, low-$\eta$ signature. Configurations 01, 03, 09, and 27 give $|V_{zz}|$ = 318.9--326.2 V/\AA$^2$ and $\eta$ = 0.055--0.082 across their Th sites (Table II and Fig. 3). This combination is close to the high-EFG site 3 reported by Hiraki et al. \cite{ref26}, rather than to the D-like or O-like benchmarks. A low $\eta$ indicates that the two transverse principal EFG components are similar; it does not imply a small EFG or a nearly spherical local charge distribution. The shared high-EFG signature thus motivates testing the possible contribution of the low-energy dimer family to site 3.

Within the sampled neutral dimer family, configurations 01, 03, 09, and 27 have the lowest relative static energies. Their EFGs can therefore be compared with the experimental high-EFG site without selecting the configurations solely for spectral agreement. In Fig. 4, separate four-site fits use the DFT EFGs of 01 and 03 for site 3, with equal contributions from the two Th sites and fixed measured frequencies \cite{ref26}. The procedure is described in Sec. VII of the Supplemental Material \cite{ref46}. The reduced $\chi^2$ values range from 8.824 to 28.898, substantially exceeding unity. Thus, the comparison tests compatibility of the splitting scale under the stated model and error assumptions, rather than demonstrating an adequate statistical fit or a unique microscopic assignment.

\begin{table}[t]
\caption{Experimental EFG parameters for $^{229}$Th in CaF$_2$ \cite{ref26}. Ranges summarize the central values reported across the measured crystals; they are not confidence intervals.}\label{tab:1}
\centering\begin{ruledtabular}
\begin{tabular}{lcc}
Experimental center \cite{ref26}  & $|V_{zz}|$ (V/\AA$^2$) & $\eta$ \\\hline
Site 1 (O-like)  & 0--2 & unresolved \\
Site 2 (D-like)  & 106--107 & 0.59--0.61 \\
Site 3 (High EFG)  & 319--320 & 0.05--0.22 \\
Site 4 (High EFG )  & 258--261 & 0.00--0.21 \\
\end{tabular}\end{ruledtabular}
\end{table}

\begin{table*}[t]
\caption{Relative static energies and signed Th-site EFGs for the lowest-energy sampled neutral dimer family. Relative energies are calculated from the total energies in Table S5 of the Supplemental Material; their displayed precision does not imply accuracy at the meV scale.}\label{tab:2}
\centering\begin{ruledtabular}
\begin{tabular}{lrrrrr}
ID & $\Delta E_{\mathrm{rel}}$ (meV) & $V_{zz}^{(1)}$ (V/\AA$^2$) & $V_{zz}^{(2)}$ (V/\AA$^2$) & $\eta_1$ & $\eta_2$ \\\hline
01 & 0.000 & -321.626 & -318.901 & 0.081 & 0.055 \\
03 & 0.480 & -321.044 & -321.044 & 0.073 & 0.073 \\
09 & 2.545 & -320.203 & -326.244 & 0.064 & 0.082 \\
27 & 3.657 & -321.555 & -321.737 & 0.065 & 0.065 \\
\end{tabular}\end{ruledtabular}
\end{table*}

Configurations 01, 03, 09, and 27 relax from different initial interstitial-F arrangements to closely similar local geometries. Their Th--Th distances span only 4.0401--4.0442 \AA (Table S7 of the Supplemental Material \cite{ref46}). Relative to 01, their static energy differences are 0.480, 2.545, and 3.657 meV for 03, 09, and 27, respectively (Table II). At fixed composition and charge, these differences are also relative formation-energy differences. Together with their similar EFG magnitudes and $\eta$ values, this near-degeneracy suggests that the four labels may represent small structural variations of a common local motif rather than four clearly distinct defect structures. This interpretation does not establish exact structural equivalence, and the meV-scale energies do not resolve their ordering or equilibrium populations.

This structural similarity is accompanied by a common EFG signature (Fig. 3). All eight Th sites have negative $V_{zz}$, with $|V_{zz}|$ = 318.9--326.2 V/\AA$^2$ and $\eta$ = 0.055--0.082. The small $\eta$ values indicate nearly equal transverse EFG components, while the large $|V_{zz}|$ reflects substantial axial anisotropy. Thus, the local EFG is approximately axially symmetric, without approaching the zero-EFG limit. The two Th sites in 03 have identical tabulated parameters, $V_{zz}$ = -321.044 V/\AA$^2$ and $\eta$ = 0.073; those in 27 are also closely matched. Configurations 01 and 09 show larger site-to-site differences in $|V_{zz}|$, of 2.725 and 6.041 V/\AA$^2$, respectively, together with small differences in $\eta$. The shared high-$|V_{zz}|$, low-$\eta$ signature is compatible with the experimental site-3 range in Table I \cite{ref26}.

The similar relative formation energies, local geometries, electric field gradients, and asymmetry parameters of configurations 01, 03, 09, and 27 suggest that they may represent closely related variants of a common local structural motif. Nevertheless, this structural similarity does not imply identical nuclear spectra. Small variations in $V_{zz}$ and $\eta$ modify the quadrupole level spacings and hence shift individual transition frequencies. If these local variants coexist and their spectral contributions remain unresolved, their superposition can lead to inhomogeneous broadening. The distinctive focus of this work is therefore the sensitivity of nuclear spectroscopic signatures to subtle local variations within a charge-neutral Th-dimer family. This provides a possible microscopic mechanism for spectral heterogeneity.

\begin{figure*}[t]\centering
\includegraphics[width=\textwidth]{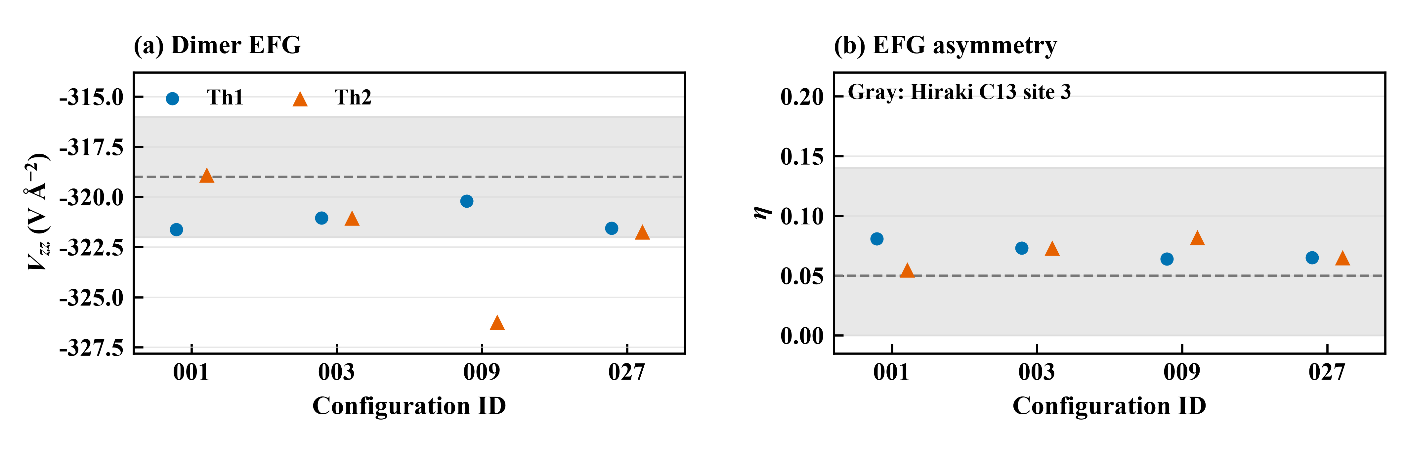}
\caption{EFG signatures of the low-energy neutral Th dimer configurations 01, 03, 09, and 27. (a) Signed $V_{zz}$. (b) Asymmetry $\eta$. Blue circles and orange triangles denote Th1 and Th2, respectively. Gray bands show the reported C13 site-3 values $V_{zz}$ = -319(3) V/\AA$^2$ and $\eta$ = 0.05(9) \cite{ref26}. Parentheses denote differences between two fitted experimental scans, not confidence intervals; the $\eta$ band is clipped at zero. Configuration labels in the figure include an additional leading zero.}\label{fig:3}
\end{figure*}

\begin{figure*}[!t]\centering
\includegraphics[width=\textwidth]{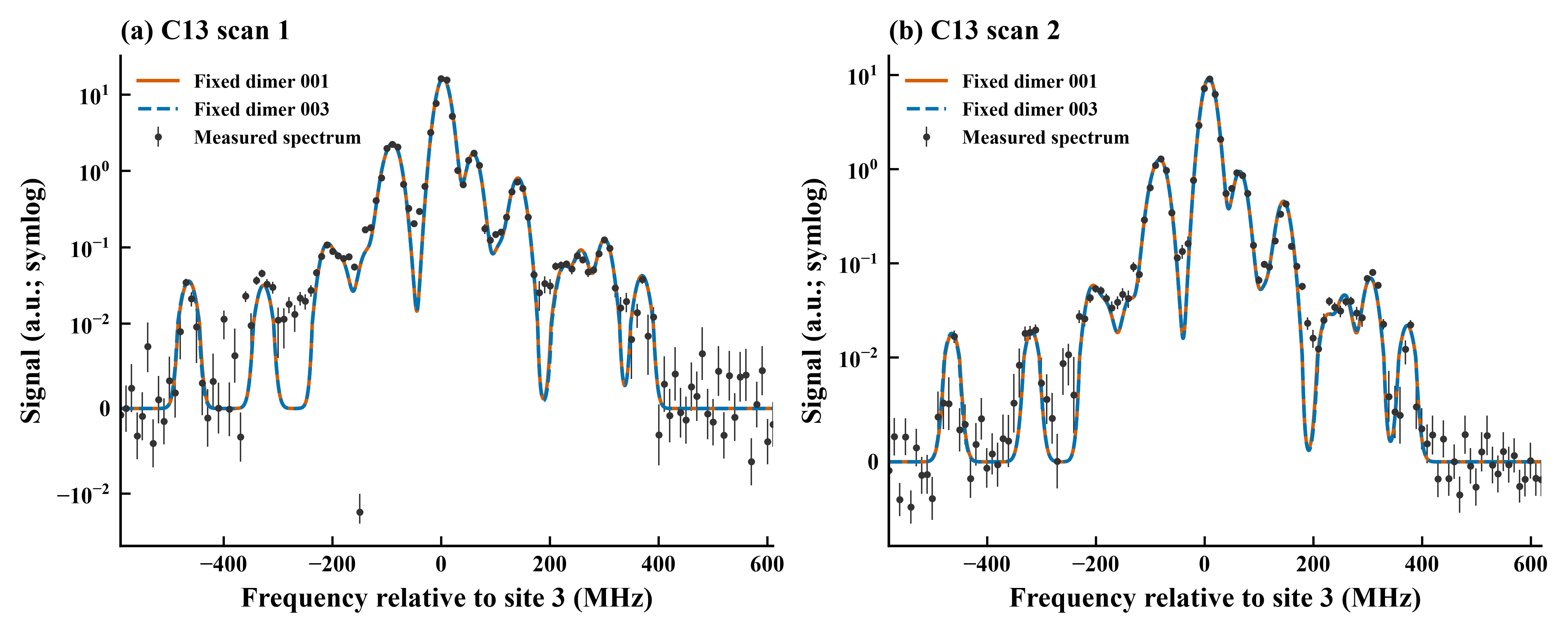}
\caption{C13 excitation spectra reported by Hiraki et al. \cite{ref26} and separate four-site fits using the fixed DFT EFGs of dimer configurations 01 (solid orange) and 03 (dashed blue) for site 3: (a) scan 1; (b) scan 2. Points show measured signals with intensity-error bars. Both Th sites contribute equally. Fits use fixed measured frequencies and intensity-error weights; each frequency axis is referenced to the site-3 center from the 01 fit. The symmetric-logarithmic ordinate retains negative signals. Detailed EFG comparisons, site-3 components, and fit parameters are provided in Sec. VII of the Supplemental Material \cite{ref46}. Configuration labels in the figure include an additional leading zero.}\label{fig:4}
\end{figure*}

Using the calculated EFGs of configurations 01 and 03, the C13 spectral comparison \cite{ref26} supports compatibility with the site-3 splitting scale (Fig. 4; see Sec. VII of the Supplemental Material \cite{ref46} for methods and parameters). The dimer EFG scales can be compared with earlier calculations, while distinguishing the compensation models. The nearest-neighbor dimer calculation without local charge compensation reported by Hiraki et al. gave $V_{zz}$ of about 95 V/\AA$^2$ and $\eta$ of about 0.6, which was proposed as a candidate for the center near 100 V/\AA$^2$ \cite{ref26}. The EFG of the neutral, four-$\mathrm{F}_{i}^{-}$-compensated configuration 30 lies in a similar range. Configurations 14 and 08 further show that $\mathrm{F}_{i}^{-}$ dimers can also obtain $V_{zz}$--$\eta$ combinations close to the D-like center. At the same time, configurations 01, 03, 09, and 27 form the low-relative-energy/high-EFG family and match the EFG scale of site 3. The fact that the same $\mathrm{F}_{i}^{-}$ dimer family contains both D-like candidate sites and high-EFG candidate sites indicates that dimers can contribute to the formation of multiple EFG environments. By contrast, the near-zero EFG O-like center has not yet been reproduced in the local  compensation models considered here. Morawetz et al. reported that the static $|V_{zz}|$ of the O-center is below 0.1 V/\AA$^2$ \cite{ref25}, and site 1 in Hiraki et al. also corresponds to $V_{zz}$ of about 0--2 V/\AA$^2$ \cite{ref26}. The smallest $|V_{zz}|$ among the local compensation models in this work is 28.1 V/\AA$^2$ (dimer-cluster 29), still higher than the near-zero EFG upper limit of the O-like center. The absence of a near-zero EFG configuration within the present model space suggests that O-like centers may originate from more symmetric compensation motifs beyond the local compensation models considered here, including remote compensation or other higher-symmetry arrangements.

\subsection{Implications for multiple spectral centers}

These comparisons suggest considerations on crystal growth and spectral control. If the goal is to obtain narrower and more reproducible nuclear-clock spectral lines, the total Th concentration alone should not be used as the optimization variable. Charge-compensation mode, local F chemical potential, Th aggregation tendency, and thermal-treatment path should also be considered. Low concentration or growth conditions favorable for remote or highly symmetric compensation may be closer to an O-like low-EFG environment. High concentration or conditions that promote nearest-neighbor Th--Th correlations may increase the weights of D-like or high-EFG dimer-cluster centers. Because the present screening does not determine chemical-potential-dependent equilibrium populations, this conclusion should be understood as a set of structural hypotheses rather than as a prediction of the dominant defect abundance under specific growth conditions.

\section{Conclusions}

This work compared the first-principles structures, relative static energies, and EFG parameters of several charge-compensating defect models in Th-doped CaF$_2$. The $\mathrm{F}_{i}^{-}$ monomer and $V_{\mathrm{Ca}}$ monomer cover part of the EFG range. In particular, the nearest-neighbor $V_{\mathrm{Ca}}$ monomer gives $|V_{zz}|$ close to the experimental D-like scale, but $\eta$ is markedly too low, indicating that the monomer models cannot fully explain the principal experimental D-like center. $\mathrm{F}_{i}^{-}$ dimer configurations span a broader EFG range and contain both D-like candidate sites and the low-relative-energy/high-EFG family. Configurations 14, 30, and 08 yield $V_{zz}$--$\eta$ combinations close to that of the D-like center, whereas configurations 01, 03, 09, and 27 form the low-relative-energy/high-EFG family and have $|V_{zz}|$ values of approximately 320 V/\AA$^2$, close to the high-EFG center observed in site-resolved experiments. The similar relative formation energies, geometries, EFGs, and $\eta$ values of configurations 01, 03, 09, and 27 suggest a common local structural motif. Small variations within this motif can nevertheless shift individual quadrupole transitions and, if physically realized and unresolved, contribute to inhomogeneous broadening. These results support Th dimers as candidate contributors to the multiple EFG environments in Th:CaF$_2$. The present local compensation models do not reproduce the near-zero-EFG O-like center. This center may correspond to a local environment with higher symmetry, remote charge compensation, or a structural mechanism distinct from the present local $2\mathrm{F}_{i}^{-}$/$V_{\mathrm{Ca}}$ models.

Materials optimization of Th:CaF$_2$ should therefore balance increased Th incorporation against preservation of a well-defined local defect environment, rather than simply maximizing total Th concentration. Studies of defect clustering in fluorite fluorides, together with recent HAADF-STEM atomic-column observations, suggest that higher Th concentrations may promote Th-related defect aggregation. The resulting local disorder and defect-related absorption or scattering could reduce VUV transmittance, while the associated diversity of Th environments may broaden the range of Th-site EFGs and contribute to spectral inhomogeneity. Future studies should combine local structural descriptors with measurements of Th concentration, VUV transmittance, and site-resolved EFG weights, while systematically controlling growth and annealing conditions and documenting the fluorine chemical-potential history within the same sample series. Correlating these data with local-structure evidence from STEM or XAFS will be essential for constraining the populations of dimer and higher-order clusters in real crystals and identifying the dominant microscopic origins of the O-like, D-like, and high-EFG centers.

\begin{acknowledgments}

We acknowledge the National Natural Science Foundation of China (12341402, 12341403, U22A20132), Strategic Priority Research Program of the Chinese Academy of Sciences (XDB0920000), and Zhangjiang Laboratory (ZJSP21A001D).\end{acknowledgments}

\section*{Data availability}

The computational data are available from the authors upon reasonable request.

\makeatletter\immediate\write\@auxout{\string\citation{prbcontrol}}\makeatother

\bibliographystyle{apsrev4-2-threeauthors}

\bibliography{references}

\end{document}

\title{Supplemental Material for ``Th$^{4+}$ ion cluster structures and electric field gradients in $^{229}$Th:CaF$_2$''}

\author{Yuxiang Li}
\thanks{These authors contributed equally to this work.}
\affiliation{Research Center of Laser Crystal, Shanghai Institute of Optics and Fine Mechanics, Chinese Academy of Sciences, Shanghai 201800, China.}
\affiliation{Center of Materials Science and Optoelectronics Engineering, University of Chinese Academy of Sciences, Beijing 100049, China.}
\author{Lin Li}
\thanks{These authors contributed equally to this work.}
\email[Contact author: ]{lilin1@siom.ac.cn}
\affiliation{Research Center of Laser Crystal, Shanghai Institute of Optics and Fine Mechanics, Chinese Academy of Sciences, Shanghai 201800, China.}
\affiliation{Center of Materials Science and Optoelectronics Engineering, University of Chinese Academy of Sciences, Beijing 100049, China.}
\author{Qiaorui Gong}
\affiliation{Research Center of Laser Crystal, Shanghai Institute of Optics and Fine Mechanics, Chinese Academy of Sciences, Shanghai 201800, China.}
\affiliation{Center of Materials Science and Optoelectronics Engineering, University of Chinese Academy of Sciences, Beijing 100049, China.}
\author{Shanming Li}
\affiliation{Research Center of Laser Crystal, Shanghai Institute of Optics and Fine Mechanics, Chinese Academy of Sciences, Shanghai 201800, China.}
\affiliation{Center of Materials Science and Optoelectronics Engineering, University of Chinese Academy of Sciences, Beijing 100049, China.}
\author{Chengchun Zhao}
\affiliation{Research Center of Laser Crystal, Shanghai Institute of Optics and Fine Mechanics, Chinese Academy of Sciences, Shanghai 201800, China.}
\affiliation{Center of Materials Science and Optoelectronics Engineering, University of Chinese Academy of Sciences, Beijing 100049, China.}
\author{Lu Deng}
\affiliation{Advanced Laser and Optoelectronic Functional Materials Department, Shanghai Institute of Optics and Fine Mechanics, Chinese Academy of Sciences, Shanghai 201800, China.}
\author{Peixiong Zhang}
\email[Contact author: ]{pxzhang@jnu.edu.cn}
\affiliation{Research Center of Laser Crystal, Shanghai Institute of Optics and Fine Mechanics, Chinese Academy of Sciences, Shanghai 201800, China.}
\affiliation{Department of Optoelectronic Engineering, Jinan University, Guangzhou, Guangdong 510632, China.}
\author{Yin Hang}
\email[Contact author: ]{yhang@siom.ac.cn}
\affiliation{Research Center of Laser Crystal, Shanghai Institute of Optics and Fine Mechanics, Chinese Academy of Sciences, Shanghai 201800, China.}
\affiliation{Center of Materials Science and Optoelectronics Engineering, University of Chinese Academy of Sciences, Beijing 100049, China.}
\author{Longsheng Ma}
\affiliation{State Key Laboratory of Precision Spectroscopy, East China Normal University, Shanghai 200062, China.}
\author{Shining Zhu}
\affiliation{National Laboratory of Solid-State Microstructures, School of Physics, Collaborative Innovation Center of Advanced Microstructures, Nanjing University, Nanjing, Jiangsu 210093, China.}

\maketitle

\tableofcontents

\clearpage

\section{Plane-wave cutoff sensitivity}

The three charged D0 controls use PBE, 96 atoms, charge +4, identical geometry, a $\Gamma$-centered 2 $\times$ 2 $\times$ 2 mesh, LREAL = .FALSE., EDIFF = $10^{-9}$ eV and no SOC. Total energies at 600, 650 and 675 eV are -736.92807576, -736.97930418 and -737.04066169 eV, respectively. The 675 eV point defines the comparison reference.

\begin{table}[h]\centering\small\caption{Cutoff sensitivity for the charged D0 geometry. The two Th sites have the same printed principal values. Energy differences are relative to the 675 eV calculation, with N = 96 atoms. Transverse components are relabeled to satisfy the principal-axis ordering in Eq. (S2); eigenvalues and $\eta$ are unchanged.}\begin{ruledtabular}\begin{tabular}{lrrrrr}

Cutoff (eV) & (E - E675)/N (meV/atom) & $V_{xx}$ (V/\AA$^2$) & $V_{yy}$ (V/\AA$^2$) & $V_{zz}$ (V/\AA$^2$) & $\eta$ \\\hline

600 & 1.172770 & -7.863 & -78.988 & 86.850 & 0.819 \\

650 & 0.639141 & -7.862 & -78.987 & 86.849 & 0.819 \\

675 & 0.000000 & -7.862 & -78.990 & 86.852 & 0.819 \\

\end{tabular}\end{ruledtabular}\end{table}

\begin{figure}[t]\centering
\includegraphics[width=0.92\textwidth]{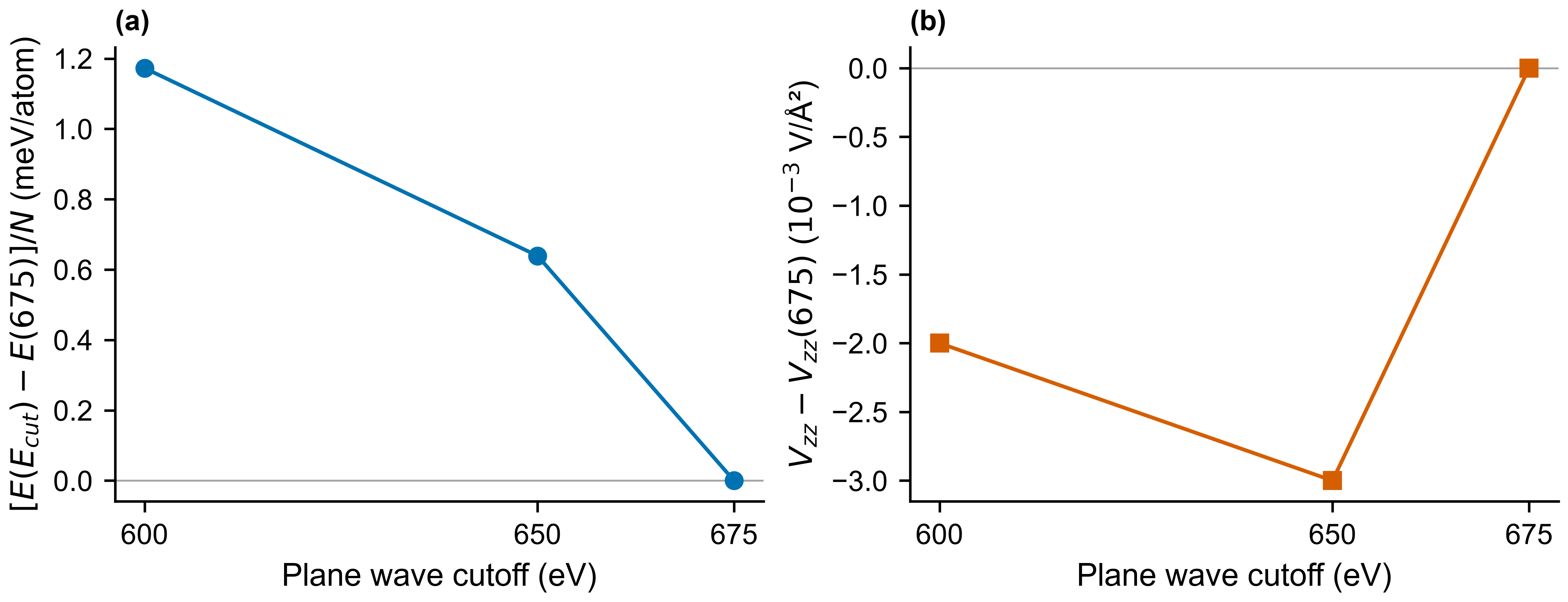}
\caption{Cutoff sensitivity at fixed D0 geometry. (a) Energy per atom relative to 675 eV, with N = 96. (b) Change of $V_{zz}$ relative to 675 eV. Connecting lines guide the eye; no extrapolated value or uncertainty band is implied.}\label{fig:s1}
\end{figure}

Over 600--675 eV, the printed $V_{zz}$ range is 0.003 V/\AA$^2$ and all $\eta$ values are 0.819. Raising the cutoff from 650 to 675 eV changes the total energy by -0.06135751 eV (-0.639141 meV/atom). These results quantify the EFG and energy sensitivity over the tested interval at one fixed geometry.

The PAW order is Ca\_sv\_GW (31Mar2010), Th (07Sep2000), F\_GW\_new (19Mar2012), distinct from the neutral screening family. Residual forces are approximately 0.223 eV/\AA; diagonal stresses vary from about -22 to -23 kbar at 600 eV to -24 to -25 kbar at 675 eV. The static controls do not establish convergence of a relaxed geometry or of stress.

All three calculations reached the electronic stopping criterion and normal termination. POSCAR/CONTCAR coordinates and lattices agree. The two Th sites have equal printed EFGs. Independent diagonalization agrees with the principal values within 0.002 V/\AA$^2$. No PSMAXN warning occurs in these controls.

\clearpage

\section{Electric field gradients and reciprocal-space sampling}

The EFG is a symmetric Cartesian tensor. Its principal values are obtained by diagonalization, and $V_{zz}$ denotes the component with the largest absolute magnitude. The sign of $V_{zz}$ is retained in the calculation tables. A principal-axis direction and its negative represent the same physical axis.

\begin{equation}V_{xx}+V_{yy}+V_{zz}=0.\end{equation}

\begin{equation}\eta=\frac{|V_{xx}-V_{yy}|}{|V_{zz}|},\qquad |V_{zz}|\geq|V_{yy}|\geq|V_{xx}|.\end{equation}

The asymmetry satisfies 0 $\leq$ $\eta$ $\leq$ 1. The absolute-value expression in Eq. (S2) is insensitive to exchanging the two transverse labels. A near-zero EFG should not be assigned a meaningful $\eta$. Printed tensor components can leave a small trace through rounding; this must be judged at the output precision.

A separate historical charged D model provides an actual k-point pair. Both calculations use the identical POSCAR and POTCAR hashes, PBE, 800 eV, 96 atoms, charge +4, ISPIN = 1, EDIFF = $10^{-9}$ eV and LREAL = .FALSE. Their INCAR parameters match apart from the descriptive SYSTEM label.

\begin{table}[h]\centering\small\caption{Matched k-point controls for the historical charged D geometry. Principal values are the same at both Th sites. Transverse components are relabeled to satisfy the principal-axis ordering in Eq. (S2); eigenvalues and $\eta$ are unchanged.}\begin{ruledtabular}\begin{tabular}{lrrrrrr}

$\Gamma$-centered mesh & Nk & Energy (eV) & $V_{xx}$ (V/\AA$^2$) & $V_{yy}$ (V/\AA$^2$) & $V_{zz}$ (V/\AA$^2$) & $\eta$ \\\hline

1 $\times$ 1 $\times$ 1 & 1 & -580.54283491 & -7.000 & -85.733 & 92.733 & 0.849 \\

2 $\times$ 2 $\times$ 2 & 8 & -580.63913619 & -6.937 & -85.842 & 92.780 & 0.850 \\

\end{tabular}\end{ruledtabular}\end{table}

Increasing the mesh from $\Gamma$ only to 2 $\times$ 2 $\times$ 2 changes E by -0.09630128 eV (-1.003138 meV/atom), $V_{zz}$ by +0.047 V/\AA$^2$ (+0.0507\%) and $\eta$ by +0.001. Both calculations reached electronic convergence and normal termination. This comparison measures sensitivity between the two meshes at fixed geometry.

\clearpage

\section{Defect models and calculation protocol}

This supplement presents calculated structures, relative static energies and electric field gradients (EFGs) for thorium defects in calcium fluoride. The neutral screening set contains 31 dimer configurations, two interstitial-compensated monomers and one Ca-vacancy-compensated monomer. Numerical controls quantify the sensitivity to plane wave cutoff, reciprocal space sampling and lattice constraints for the specified models.

\begin{table}[h]\centering\small\caption{Composition and charge compensation of the model families. Atom counts refer to the 32-cation-site parent supercell.}\begin{ruledtabular}\begin{tabular}{p{0.19\textwidth}p{0.2\textwidth}p{0.08\textwidth}p{0.45\textwidth}}

Model & Composition & Atoms & Compensation \\\hline

Neutral monomer M & Ca$_{31}$ThF$_{66}$ & 98 & One Th substitution and two F interstitials \\

Neutral dimer C & Ca$_{30}$Th$_2$F$_{68}$ & 100 & Two Th substitutions and four F interstitials \\

Vacancy monomer & Ca$_{30}$ThF$_{64}$ & 95 & One Th substitution and one Ca vacancy \\

Charged D control & Ca$_{30}$Th$_2$F$_{64}$ & 96 & Net charge +4 without local compensation \\

\end{tabular}\end{ruledtabular}\end{table}

Calculations used VASP with GGA-PBE and PAW potentials in the Ca\_sv--Th--F order. Table S4 lists the relaxation and static EFG settings. Relaxations reaching the 120--200-step limit were continued until the force criterion was met. Static calculations used a 700 eV cutoff and retained the relaxation k-point mesh and PAW potentials. Th-site EFG tensors were diagonalized as described in Sec. II; the nuclear quadrupole moments and spectral model used to convert EFGs to quadrupole splittings are specified in Sec. VII.

\begin{table}[h]\centering\small\caption{Parameters of the original scalar-relativistic screening series. Static ENCUT is corrected to 700 eV using the archived screening provenance, consistently with Sec. VII.}\begin{ruledtabular}\begin{tabular}{p{0.28\textwidth}p{0.3\textwidth}p{0.3\textwidth}}

Parameter & Relaxation & Static EFG \\\hline

ENCUT (eV) & 650 & 700 \\

KPOINTS & Gamma 3 $\times$ 3 $\times$ 3 & Gamma 3 $\times$ 3 $\times$ 3 \\

EDIFF (eV) & $10^{-6}$ & $10^{-8}$ \\

EDIFFG (eV/\AA) & -0.02 & not applicable \\

IBRION & 2 & -1 \\

ISIF & 3 & 2 \\

NSW & 120--200 & 0 \\

ISYM & 0 & 0 \\

ISPIN & 1 & 1 \\

LEFG & not required & .TRUE. \\

LREAL & per input archive & .FALSE. \\

LASPH & per input archive & .TRUE. \\

ADDGRID & per input archive & .TRUE. \\

\end{tabular}\end{ruledtabular}\end{table}

The screening series uses scalar-relativistic PBE. The energies and EFGs are screening results at the stated settings. The controls in Sections I, II, and VI use separately specified PAW data and cell constraints. The energy comparison is restricted to relative static energies within a fixed-composition, fixed-charge family. It does not determine aggregation energies or equilibrium populations.

\clearpage

\section{Dimer screening energies and EFGs}

The energy zero is the minimum static total energy within the Ca$_{30}$Th$_2$F$_{68}$ family. Each configuration contains two Th sites. Signed $V_{zz}$ values are retained; the displayed precision is not an estimate of absolute accuracy.

\begin{equation}\Delta E_{\mathrm{rel},i}=E_{\mathrm{tot}}(C_i)-E_{\mathrm{tot}}(C_{01}).\end{equation}

\begin{table}[h]\centering\small\caption{Complete dimer screening records. E and $\Delta$E are in eV; $V_{zz}$ is in V/\AA$^2$; $\eta$ is dimensionless.}\begin{ruledtabular}\begin{tabular}{lrrrrrr}

ID & $E_{\mathrm{tot}}$ & $\Delta E_{\mathrm{rel}}$ & $V_{zz}$ Th1 & $\eta$ Th1 & $V_{zz}$ Th2 & $\eta$ Th2 \\\hline

01 & -589.67924079 & 0.000000 & -321.626 & 0.081 & -318.901 & 0.055 \\

02 & -587.86538471 & 1.813856 & -82.222 & 0.641 & -82.222 & 0.641 \\

03 & -589.67876063 & 0.000480 & -321.044 & 0.073 & -321.044 & 0.073 \\

04 & -589.12388088 & 0.555360 & -244.202 & 0.972 & -218.053 & 0.795 \\

05 & -589.44470180 & 0.234539 & -310.672 & 0.575 & -119.834 & 0.315 \\

06 & -589.43284775 & 0.246393 & -284.948 & 0.750 & -72.640 & 0.819 \\

07 & -588.05243923 & 1.626802 & -281.752 & 0.759 & -281.752 & 0.759 \\

08 & -588.70159440 & 0.977646 & 100.303 & 0.666 & 100.303 & 0.666 \\

09 & -589.67669541 & 0.002545 & -320.203 & 0.064 & -326.244 & 0.082 \\

10 & -589.23439661 & 0.444844 & 193.769 & 0.936 & -103.137 & 0.786 \\

11 & -588.94880768 & 0.730433 & 240.920 & 0.344 & 219.647 & 0.144 \\

12 & -588.01491074 & 1.664330 & 224.783 & 0.647 & -356.436 & 0.158 \\

13 & -589.43536057 & 0.243880 & -287.528 & 0.768 & -72.770 & 0.530 \\

14 & -589.23569887 & 0.443542 & -204.079 & 0.961 & -103.543 & 0.652 \\

15 & -589.45563945 & 0.223601 & -224.703 & 0.559 & 215.579 & 0.761 \\

16 & -589.46791100 & 0.211330 & -128.734 & 0.987 & 164.174 & 0.829 \\

17 & -589.43030928 & 0.248932 & -280.841 & 0.796 & 75.000 & 0.958 \\

18 & -589.22816776 & 0.451073 & 215.574 & 0.609 & -102.742 & 0.733 \\

19 & -589.42570530 & 0.253535 & 81.740 & 0.903 & -278.886 & 0.754 \\

20 & -589.10545938 & 0.573781 & -163.307 & 0.441 & -116.626 & 0.929 \\

21 & -588.44756302 & 1.231678 & 230.574 & 0.450 & -272.820 & 0.377 \\

22 & -587.86726389 & 1.811977 & -241.186 & 0.113 & 352.881 & 0.762 \\

23 & -587.09186186 & 2.587379 & -344.746 & 0.253 & -344.746 & 0.253 \\

24 & -588.12168111 & 1.557560 & 146.173 & 0.661 & -391.203 & 0.159 \\

25 & -589.10677090 & 0.572470 & -166.563 & 0.410 & 132.482 & 0.874 \\

26 & -588.89113656 & 0.788104 & 73.948 & 0.748 & 73.948 & 0.748 \\

27 & -589.67558409 & 0.003657 & -321.555 & 0.065 & -321.737 & 0.065 \\

28 & -589.04075212 & 0.638489 & 40.058 & 0.273 & 40.058 & 0.273 \\

29 & -589.04579464 & 0.633446 & 28.090 & 0.705 & 28.091 & 0.705 \\

30 & -589.15321482 & 0.526026 & 95.767 & 0.637 & 95.767 & 0.637 \\

31 & -588.96454753 & 0.714693 & 176.932 & 0.523 & 176.932 & 0.523 \\

\end{tabular}\end{ruledtabular}\end{table}

The sampled relative energies span 0--2.587379 eV. All 31 records are retained without selection by energy or EFG. The representative output checks described in Section III do not uniformly revalidate every record. Differences at the meV scale are not resolved as a robust energetic ordering.

\clearpage

\section{Monomer energetics and selected dimer geometries}

Monomer 01 and monomer 02 have the same composition and therefore share a meaningful relative-energy reference. The Ca-vacancy monomer has a different stoichiometry, so its total energy is not ranked against the interstitial-compensated monomers without a reservoir definition.

\begin{table}[h]\centering\small\caption{Monomer screening energies and EFG descriptors. The vacancy row has no assigned cross-composition relative energy.}\begin{ruledtabular}\begin{tabular}{lrrrr}

Defect & $E_{\mathrm{tot}}$ (eV) & $\Delta E_{\mathrm{rel}}$ (eV) & $V_{zz}$ (V/\AA$^2$) & $\eta$ \\\hline

$V_{\mathrm{Ca}}$ monomer & -554.433403 & N/A & -102.499 & 0.118 \\

$\mathrm{F}_{i}^{-}$ monomer 01 & -572.007629 & 0.000000 & 151.004 & 0.112 \\

$\mathrm{F}_{i}^{-}$ monomer 02 & -571.196558 & 0.811071 & -378.300 & 0.000 \\

\end{tabular}\end{ruledtabular}\end{table}

Table S7 lists geometric descriptors of the relaxed cells corresponding to the dimer screening records. The shortest Th--F distance is the minimum over all periodic fluorine neighbors. It is not a shell average. Configurations 01, 03, 09 and 27 have similar local distances and static energy separations below 0.004 eV. At fixed composition and charge, the relative static energies also represent relative formation energies. Their near-degeneracy, similar local geometries, and similar EFGs and $\eta$ values suggest that 01, 03, 09, and 27 may be closely related realizations of a common local motif. Their small site-specific EFG differences can nevertheless shift quadrupole line positions. This does not establish four distinct thermodynamic species or their coexistence; the possible contribution of unresolved local variations to inhomogeneous broadening is discussed in the main text.

\begin{table}[h]\centering\small\caption{Selected relaxed dimer geometries and within-family relative energies.}\begin{ruledtabular}\begin{tabular}{lrrrr}

ID & $\Delta E_{\mathrm{rel}}$ (eV) & Th--Th (\AA) & Shortest Th--F (\AA) & Cell volume (\AA$^3$) \\\hline

01 & 0.000000 & 4.0412 & 2.3029 & 1379.884 \\

03 & 0.000480 & 4.0407 & 2.3034 & 1379.903 \\

09 & 0.002545 & 4.0420 & 2.3031 & 1380.160 \\

27 & 0.003657 & 4.0401 & 2.3029 & 1379.140 \\

14 & 0.443542 & 3.9540 & 2.2943 & 1383.708 \\

30 & 0.526026 & 4.2590 & 2.2945 & 1381.470 \\

08 & 0.977646 & 4.1329 & 2.2939 & 1385.509 \\

28 & 0.638489 & 4.2594 & 2.2882 & 1381.966 \\

\end{tabular}\end{ruledtabular}\end{table}

\clearpage

\section{Fixed- and relaxed-lattice comparison}

The completed 98-atom neutral monomer control starts from the fixed-cell relaxed structure and releases the lattice through ISIF = 3. Both final static calculations use PBE, 800 eV, $\Gamma$-centered 3 $\times$ 3 $\times$ 3 sampling, EDIFF = $10^{-8}$ eV, LREAL = .FALSE., ISPIN = 1 and no SOC. The archived preparation record confirms the same Ca\_sv/Th/F PAW hash and unchanged static parameters.

\begin{table}[h]\centering\small\caption{Mechanical-boundary sensitivity of the same 98-atom neutral defect. $\mathrm{F}_{i}^{-}$ denotes the two originally appended atoms, indices 97 and 98. Transverse labels follow Eq. (S2). The static-force unit is eV/\AA.}\begin{ruledtabular}\begin{tabular}{p{0.42\textwidth}p{0.25\textwidth}p{0.25\textwidth}}

Quantity & Fixed lattice & Relaxed lattice \\\hline

Relaxation ISIF & 2 & 3 \\

a / \AA & 11.02456227 & 11.08231351 \\

b / \AA & 11.02456227 & 11.08231351 \\

c / \AA & 11.02456227 & 11.12189678 \\

Volume / \AA$^3$ & 1339.936028 & 1365.948868 \\

Static total energy / eV & -571.90503818 & -572.07592414 \\

$V_{yy}$ / V \AA$^{-2}$ & -104.359 & -84.313 \\

$V_{xx}$ / V \AA$^{-2}$ & -52.953 & -61.278 \\

$V_{zz}$ / V \AA$^{-2}$ & 157.311 & 145.591 \\

$\eta$ & 0.327 & 0.158 \\

Maximum ionic force / eV \AA$^{-1}$ & 0.0037086 & 0.0046552 \\

Maximum static force / eV \AA$^{-1}$ & 0.0037650 & 0.0046778 \\

Th--$\mathrm{F}_{i}^{-}$ distances / \AA & 2.298093, 2.298093 & 2.298700, 2.298700 \\

$\mathrm{F}_{i}^{-}$--Th--$\mathrm{F}_{i}^{-}$ angle / degrees & 70.798648 & 70.831114 \\

\end{tabular}\end{ruledtabular}\end{table}

Releasing the lattice increases the volume by 1.941349\% and decreases the static total energy by 0.17088596 eV. $V_{zz}$ changes from 157.311 to 145.591 V/\AA$^2$ (-7.450\%), and $\eta$ changes from 0.327 to 0.158. Thus, the lattice constraint has a measurable effect on the calculated EFG for this finite cell.

The free-cell relaxation completed 24 ionic steps and met the 0.005 eV/\AA force criterion; the final static calculation terminated normally. The final relaxation stress components are (0.02017, 0.02017, -0.07827, -0.03403, 0, 0) kbar in xx, yy, zz, xy, yz, zx order. The cell includes shear ($\gamma$ = 89.717430$^\circ$), and volumes are lattice determinants. Static structures and independently diagonalized EFG tensors agree with the recorded results at printed precision.

This comparison isolates a change of mechanical boundary condition within the stated 98-atom model. Its volume reference is the fixed defect cell. It does not quantify a dilute-limit or supercell-size error. The fixed-cell result retains the Th PAW PSMAXN warning; a complete warning inventory is unavailable for the free-cell output.

\clearpage

\section{C13 nuclear spectra and dimer comparison}

We compare the locally compensated 2Th + 4$\mathrm{F}_{i}^{-}$ dimer with site 3 in the C13 crystal measured by Hiraki et al. \cite{ref26}. Both released C13 scans are fitted independently, using all 121 points per scan. The measured frequencies are fixed, and only the reported intensity errors are used as fit weights. Figure S2 compares the calculated electric field gradients (EFGs) with the C13 reference and shows the full spectra and fitted site-3 components. The comparison tests compatibility of the splitting scale; it does not establish a unique microscopic assignment. The data files used here are spectrum\_C13\_1.csv and spectrum\_C13\_2.csv.

Only configurations 01 and 03 from Table S5 are used in this spectral comparison. Configuration 01 has the lowest screening energy and is the plotting reference. Its two Th sites have ($V_{zz}$, $\eta$) = (-321.626 V/\AA$^2$, 0.081) and (-318.901 V/\AA$^2$, 0.055); configuration 03 has (-321.044 V/\AA$^2$, 0.073) at both Th sites. These are the original 100-atom neutral Ca$_{30}$Th$_2$F$_{68}$ screening results in Table S5, obtained with 650 eV relaxation and 700 eV static calculations, PBE, ISPIN = 1, and no spin--orbit coupling.

For each EFG, we diagonalize the nuclear electric quadrupole Hamiltonian in its principal-axis frame, with dimensionless spin operators:

\begin{equation}\frac{\hat H_Q}{h}=\frac{C Q V_{zz}}{4I(2I-1)}\left[3\hat I_z^2-I(I+1)+\frac{\eta}{2}(\hat I_+^2+\hat I_-^2)\right].\end{equation}

Here C = 2.417989242 MHz/(eb · V/\AA$^2$), Q is expressed in eb, and $V_{zz}$ is in V/\AA$^2$. We use ground and excited spins $I_g$ = 5/2 and $I_e$ = 3/2 and $Q_g$ = 3.11 eb; $Q_e$/$Q_g$ is fitted within 0.53--0.61. Magnetic-dipole branch strengths are calculated from the mixed eigenvectors and Clebsch--Gordan coefficients, summed over polarization components and Kramers partners, and normalized over the six branches. Equal ground-state populations and an isotropic orientation average are assumed; crystal-specific polarization and direct nuclear coupling are omitted.

The four-site spectrum uses one Gaussian for unresolved site 1 and quadrupole-split components for sites 2--4. For each configuration separately, site 3 is the equally weighted sum of the two calculated Th spectra, with a common center and amplitude and fixed DFT EFGs. Sites 2 and 4 each have adjustable $V_{zz}$, $\eta$, center, and amplitude; site 1 has an adjustable center and amplitude. One Gaussian width and one $Q_e$/$Q_g$ ratio are shared within each scan. Each fit has 14 spectral parameters and no additive background.

We minimize the sum of squared differences between measured and fitted intensities divided by their reported variances, evaluated at the measured frequencies. Each frequency remains fixed throughout the fit. Frequency uncertainties, latent frequency shifts, and frequency-shift penalties are not included.

There are N = 121 measured intensities and k = 14 spectral parameters per fit, giving N - k = 107 degrees of freedom. All measured points, including negative signals, are retained. We assume independent intensity errors and do not rescale them to force a unit reduced $\chi^2$. This conditional fit does not account for the effect of frequency uncertainty on goodness of fit or parameter errors.

\begin{table}[h]\centering\small\caption{Parameters from the fixed-EFG fits for configurations 01 and 03, using intensity-error weights and fixed measured frequencies. Each configuration is fitted separately to both C13 scans. }\begin{ruledtabular}\begin{tabular}{p{0.4\textwidth}p{0.135\textwidth}p{0.135\textwidth}p{0.135\textwidth}p{0.135\textwidth}}

Quantity & 01 Scan 1 & 01 Scan 2 & 03 Scan 1 & 03 Scan 2 \\\hline

Reduced $\chi^2$ & 28.810 & 8.838 & 28.898 & 8.824 \\

Gaussian FWHM / MHz & 28.79 & 29.50 & 28.79 & 29.50 \\

Site-3 signal fraction / \% & 1.231 & 1.074 & 1.219 & 1.076 \\

Site-3 - site-2 center / MHz & +1.80 & -4.12 & +2.05 & -3.86 \\

$Q_e$ / $Q_g$ & 0.571753 & 0.572007 & 0.571785 & 0.571997 \\

\end{tabular}\end{ruledtabular}\end{table}

\begin{table}[h]\centering\small\caption{Intensity-only fit quality for configurations 01 and 003. Entries are $\chi^2$ / degrees of freedom, evaluated at the fixed measured frequencies with the reported intensity errors. Each fit uses 121 observations and 14 spectral parameters.}\begin{ruledtabular}\begin{tabular}{lrrr}

Site-3 model & Parameters k & C13 scan 1 & C13 scan 2 \\\hline

Fixed dimer 01 & 14 & 3082.70 / 107 & 945.70 / 107 \\

Fixed dimer 03 & 14 & 3092.06 / 107 & 944.16 / 107 \\

\end{tabular}\end{ruledtabular}\end{table}

The reported C13 site-3 reference is $V_{zz}$ = -319(3) V/\AA$^2$ and $\eta$ = 0.05(9). The parentheses quantify differences between the two fitted scans, not standard errors or confidence intervals. Figure S2 clips the illustrative $\eta$ band at zero. Agreement with these reference values supports compatibility of the local EFG scale, while the scan dependence, fit quality, and intensity assumptions limit structural inference. The equal 1:1 Th weighting is a model assumption and fitted signal fractions cannot be converted into defect populations without excitation and detection factors. The reduced $\chi^2$ values of 8.824--28.898 are substantially larger than unity; the model is not statistically adequate under the stated independent intensity-error model. Compatibility of an EFG scale should therefore not be described as a unique or statistically validated structural assignment.

\begin{figure}[t]\centering
\includegraphics[width=0.84\textwidth]{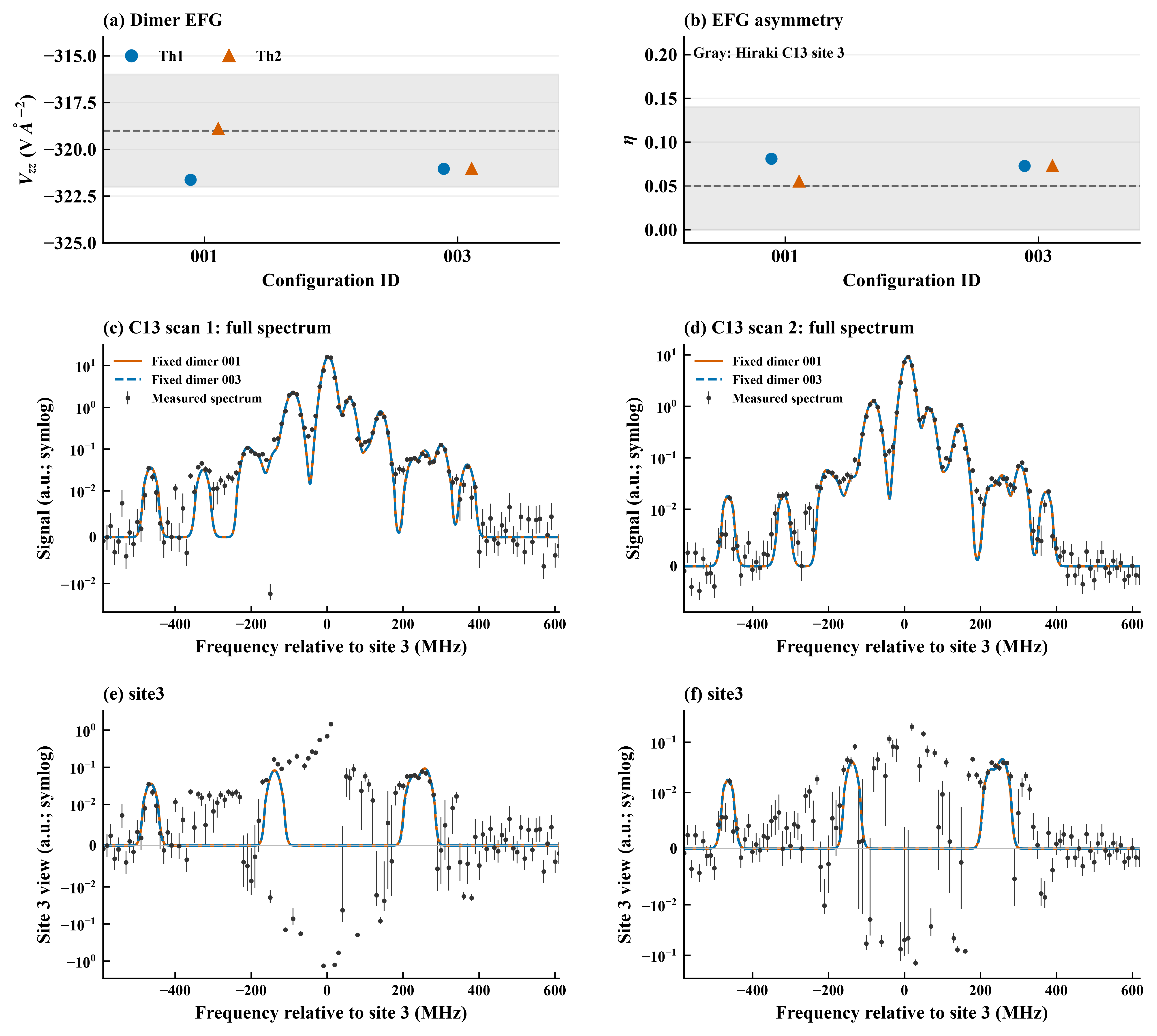}
\caption{Electric field gradients and C13 spectral fits using dimer configurations 01 and 003. (a,b) Signed $V_{zz}$ and asymmetry $\eta$ for both Th sites. Gray bands show the reported C13 site-3 values $V_{zz}$ = -319(3) V/\AA$^2$ and $\eta$ = 0.05(9); parentheses denote differences between two fitted scans, not confidence intervals, and the $\eta$ band is clipped at zero. (c,d) Measured spectra and separate four-site fits with the calculated site-3 EFGs fixed: solid orange, 01; dashed blue, 003. Both Th sites contribute equally. (e,f) Fitted site-3 components; points show observations minus sites 1, 2, and 4 from the 01 fit.  Configuration labels in the figure include an additional leading zero.}\label{fig:s2}
\end{figure}

\makeatletter\immediate\write\@auxout{\string\citation{prbcontrol}}\makeatother

\bibliographystyle{apsrev4-2-threeauthors}

\bibliography{references}